\documentclass[lettersize,journal]{IEEEtran}
\usepackage{amsmath,amsfonts}
\usepackage{algorithmic}
\usepackage{algorithm}
\usepackage{array}
\usepackage[caption=false,font=normalsize,labelfont=sf,textfont=sf]{subfig}
\usepackage{textcomp}
\usepackage{stfloats}
\usepackage{url}
\usepackage{verbatim}
\usepackage{graphicx}
\usepackage{cite}
\usepackage{amssymb}
\usepackage{xcolor}
\usepackage{siunitx}
\usepackage{multirow}
\usepackage{tabularx}
\usepackage[hidelinks]{hyperref}
\usepackage{cleveref}

\begin{document}

\title{HAWC-Leak: Heterogeneous Architecture Workload Classification utilizing Side-Channel Leakage}

\author{
Muhtasim Alam Chowdhury,
Arthur Kim,
Rozhin Yasaei
and Soheil Salehi
\vspace{-1 em}%
\thanks{Muhtasim Alam Chowdhury, Arthur Kim, and Soheil Salehi are with the Department of Electrical and Computer Engineering, University of Arizona, Tucson, AZ 85721 USA (e-mail: mmc7@arizona.edu; akim4@arizona.edu; ssalehi@arizona.edu).}
\thanks{Rozhin Yasaei is with the College of Information Science, University of Arizona, Tucson, AZ 85721 USA (e-mail: yasaei@arizona.edu).}
\thanks{Corresponding author: Muhtasim Alam Chowdhury (email: mmc7@arizona.edu).}
\thanks{Manuscript received October, 2026.}}


{

\maketitle
\begin{abstract}
Modern heterogeneous computing platforms, which integrate general-purpose processors with specialized accelerators such as FPGAs, have become the standard for high-performance and embedded applications. However, the shared physical resources of these systems introduce critical security vulnerabilities, where power consumption patterns can leak sensitive runtime information. While side-channel analysis is well established for extracting cryptographic keys from isolated algorithms, existing methods struggle to profile complex, dynamic systems where diverse workloads coexist on shared hardware resources. This paper presents HAWC-Leak, a dynamic workload classification framework designed to identify specific hardware accelerator tasks in real time using non-invasive power analysis. Unlike prior studies that focus on extracting cryptographic secrets from isolated implementations, our approach targets the runtime behavior of heterogeneous systems, leveraging machine learning to predict the active workload from feature-extracted power traces captured during execution. We implement a complete hardware-in-the-loop measurement testbed using the ChipWhisperer CW305 Artix-7 FPGA platform to capture high-resolution power signatures of diverse workloads, including Advanced Encryption Standard (AES-128), Elliptic Curve Cryptography (ECC P-256), Secure Hash Algorithm (SHA-256), Convolutional Neural Network (CNN), and a system idle state. To address the challenge of distinguishing these tasks under realistic conditions, we develop a Hierarchical Random Forest classifier that processes real-time power data to identify the executing application in controlled mixed-activity scenarios where multiple accelerators are instantiated, however, only one executes at a time. Experimental results demonstrate that this framework achieves 100\% classification accuracy across all evaluated scenarios, confirming that power consumption patterns serve as a reliable and distinctive fingerprint for profiling heterogeneous runtime behavior. 
\end{abstract}

\begin{IEEEkeywords}
Hardware Security, Heterogeneous Architecture, Machine Learning, Side-channel Analysis.
\end{IEEEkeywords}

\section{Introduction}
\IEEEPARstart{M}{odern} computing systems are increasingly heterogeneous, integrating general-purpose CPUs with GPUs, FPGAs, and domain-specific AI accelerators to meet the escalating performance and energy-efficiency demands of emerging workloads. These platforms are now the backbone of time-critical and security-sensitive applications, ranging from autonomous driving to edge-based inference and cryptographic offloading \cite{PECCERILLO2022102561, 9530260, 8653908, 10.1145/3769087}. By dynamically scheduling tasks across these diverse processing elements, these systems achieve performance per watt that traditional architectures cannot match\cite{9530260}. 
While logical isolation mechanisms like virtualization and Trusted Execution Environments (TEEs) are designed to separate workloads, they cannot mask the physical signatures of hardware execution~\cite{8835233}. Every computation, whether a neural network layer or an encryption round, consumes power in a pattern distinct to its algorithm. Side-Channel Analysis (SCA) has traditionally been viewed through the lens of cryptanalysis, where the goal is to extract a specific secret key from a single, isolated algorithm \cite{10.1007/3-540-48405-1_25, 8835233, 9499833, 7011392}. However, less attention has been paid to a broader and potentially more dangerous threat, using these physical signals to profile the system behavior in real-time. 
If an attacker can identify exactly when and how different computational workloads execute, they can infer proprietary algorithms, monitor user activity, or launch targeted fault injection attacks during critical processing windows~\cite{10.1145/2382196.2382230, Chen2015ScheduleBasedSA, 10122602}. 

Much of this literature focuses on isolated accelerators or single-purpose devices, rather than realistic heterogeneous systems where multiple workloads coexist and interact at runtime. A key challenge in realizing this threat model on real hardware is the presence of heterogeneous execution noise, arising from concurrent accelerator activity and system-level background processes. Unlike a controlled lab environment where a target device runs a single loop, a real heterogeneous system executes mixed workloads simultaneously. A short, bursty AES encryption might be interleaved with a long-running Neural Network inference or background operating system noise~\cite{10.1109/TVLSI.2020.3027711}. Existing detection methods often fail in these scenarios because they rely on simple thresholding or non-hierarchical classification that cannot distinguish between the sharp spectral signature of a cipher and the broad power envelope of an AI accelerator. This gap necessitates a more robust, ML-driven approach capable of disentangling these overlapping signals.

In this work, we focus specifically on real-time power side-channel leakage as a practical and high-fidelity signal for workload identification in heterogeneous systems. Power consumption directly reflects switching activity within accelerator logic and is tightly coupled to both algorithm structure and execution intensity. Importantly, measurement platforms such as ChipWhisperer enable fine-grained, synchronized power-trace acquisition from FPGA fabrics, making real-time observation feasible in adversarial settings~\cite{10.1007/978-3-319-10175-0_17}. Despite this capability, a gap remains in understanding how real-time power traces can be leveraged to identify diverse workloads executing within a heterogeneous, accelerator-rich system.

\begin{figure}[!t]
    \centering
    \includegraphics[width=0.65\linewidth]{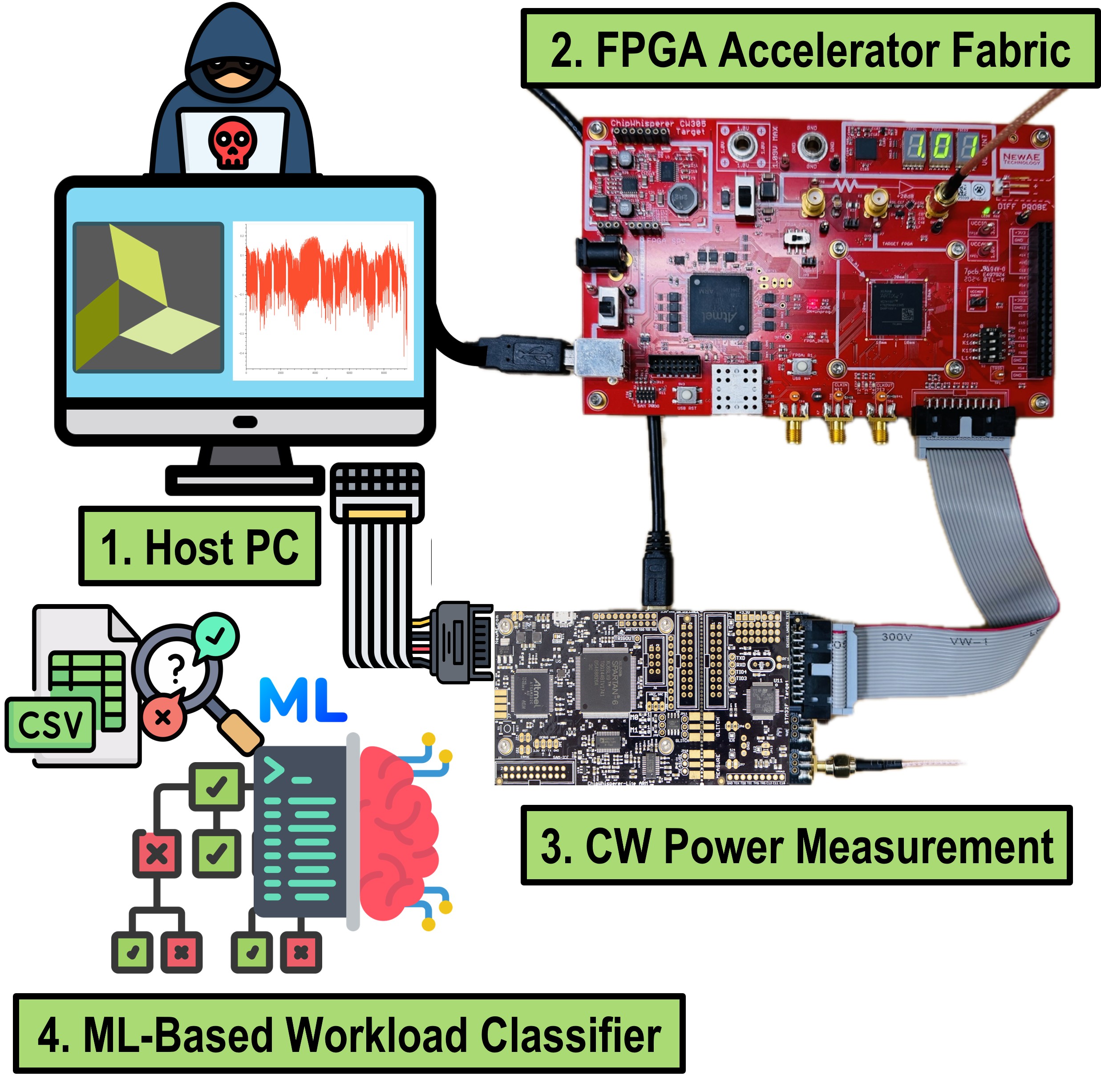}
    \caption{Overview of the HAWC-Leak framework showing FPGA execution, ChipWhisperer (CW) power capture, and ML-based workload classification end-to-end pipeline.}
    \label{fig:dynsca_overview}
    \vspace{-1 em}
\end{figure}

This paper addresses that gap by presenting HAWC-Leak, a power-trace driven framework for dynamic workload classification in heterogeneous architectures. To validate this approach, we constructed a complete end-to-end measurement pipeline using the ChipWhisperer (CW) platform and a Xilinx Artix-7 FPGA.  We emulated a realistic edge-computing scenario by deploying a suite of diverse IP cores, including AES-128, ECC P-256, SHA-256, and a Convolutional Neural Network (CNN)~\cite{Badhan2025CNNVerilog, HERON20098, Secworks_sha256_2025}. In addition to isolated executions, we generate controlled mixed-activity power traces in which multiple accelerators are instantiated on the same fabric, while only one executes at a time, introducing background leakage from inactive but clocked modules. These conditions reflect realistic deployment environments, where overlapping static and dynamic power components obscure individual workload signatures. Our results demonstrate that while traditional classifiers flounder with mixed signals, our hierarchical model achieves near-perfect accuracy in identifying the active workload. This confirms that power consumption acts as a high-fidelity barcode for hardware execution, one that remains readable even in the noisy environment of a modern heterogeneous architecture. Figure~\ref{fig:dynsca_overview} illustrates the end-to-end HAWC-Leak pipeline, from power capture to workload classification. The main contributions of this paper are summarized as follows:
\begin{itemize}
    \item We present a real-time power side-channel analysis framework for heterogeneous systems, targeting FPGA-based accelerators under realistic execution conditions.
    \item We demonstrate that diverse workloads, including cryptographic and AI accelerators, exhibit distinct and learnable power signatures, even when multiple applications coexist on the same fabric.
    \item We design and evaluate an ML-driven hierarchical classifier capable of accurately identifying active workloads from raw power traces.
\end{itemize}

The remainder of this paper is organized as follows: Section~II provides background on side-channel analysis, ChipWhisperer-based power measurement, and ML-driven detection. Section~III presents the experimental methodology, followed by Section~IV, which discusses the results. Finally, Section~V outlines future research directions, and Section~VI concludes the paper.

\section{RELATED WORKS}
Motivated by the exposure of power-based workload leakage in heterogeneous platforms, this section reviews relevant prior work. We cover heterogeneous side-channel analysis, FPGA-based power measurement, ML workload profiling, and the threat model.

\vspace{-0.7 em}
\subsection{Side-Channel Analysis in Heterogeneous Systems}
Early formalization of power-based side-channel analysis is commonly attributed to Kocher et al. in~\cite{Kocher2011IntroductionTD, 10.1007/3-540-48405-1_25}, whose work on Differential Power Analysis (DPA) demonstrated that instantaneous power consumption can be correlated with intermediate data values processed by a device
While these classical techniques effectively extracted cryptographic secrets from simple, isolated execution environments like smart cards, they typically relied on the assumption of a noise-free, synchronous acquisition path.
Furthermore, leakage is often distributed across multiple stages of computation rather than confined to a single operation, reinforcing the need for analysis techniques that capture global characteristics of power traces~\cite{8432085}.
However, the paradigm shift toward heterogeneous computing has invalidated many of these assumptions, introducing complex, system-level attack surfaces that extend beyond the boundaries of a single cryptographic core.

In modern heterogeneous architectures, the tight coupling of general-purpose processors with specialized accelerators (FPGAs, GPUs) creates new, cross-component leakage vectors. Zhao et al. in~\cite{8418606} demonstrated that FPGAs are susceptible to remote power SCAs, showing that on-chip voltage sensors can be exploited to recover keys without physical access, effectively turning the reconfigurable fabric into a sensor against itself. 
Recent work has further demonstrated covert and efficient cross-domain attacks in FPGA cloud environments, where shared on-chip resources can be exploited to extract sensitive information without direct physical access~\cite{10923723}.
Similarly, Dutta et al. revealed in “Leaky Buddies" that integrated CPU-GPU systems suffer from significant information leakage due to contention on shared resources such as the ring bus and Last Level Cache (LLC), allowing unprivileged processes to infer the activity of co-located workloads~\cite{9499833}. Despite these vulnerabilities, accurate workload fingerprinting in these environments remains an unsolved challenge due to ``heterogeneous noise''. As noted by Ghanim et al. in the analysis of the HopperFish framework, the dynamic behavior of runtime schedulers and the interleaving of diverse tasks introduce substantial variance in execution signatures~\cite{10.1145/3769087}. This necessitates a move beyond simple statistical correlation toward robust, data-driven classification methods capable of disentangling target signals from the chaotic background noise of a unified heterogeneous runtime.

\vspace{-0.7 em}
\subsection{FPGA-Based Power Measurement}
To ensure experimental reproducibility and high-fidelity signal acquisition, we leverage the ChipWhisperer framework, an open-source platform specifically designed for embedded hardware security evaluation~\cite{10.1007/978-3-319-10175-0_17}. Unlike traditional oscilloscope-based setups that suffer from clock drift, the ChipWhisperer architecture employs synchronous sampling, where the Analog-to-Digital Converter (ADC) is phase-locked to the target device's operating clock. This synchronization significantly reduces timing jitter and the required sample rate, enabling the precise capture of the high-frequency transient signals inherent to FPGA logic execution without significantly reducing the need for trace averaging. Utilizing the CW305 Artix-7 A100T target board, we emulate a realistic heterogeneous accelerator fabric while maintaining direct, low-noise access to the specific power rails of the FPGA core. This methodology is widely adopted as a practical proxy for analyzing microarchitectural power leakage, as direct FPGA rail access exposes switching-level activity typically masked by power management circuitry. 

\vspace{-0.7 em}
\subsection{Machine Learning for Workload Profiling}
The integration of ML and Deep Learning (DL) into SCA has significantly advanced the field beyond traditional statistical methods like Correlation Power Analysis (CPA).
Recent deep learning-based side-channel approaches have demonstrated strong resilience to noise and constrained measurement conditions, enabling effective leakage exploitation even in imperfect acquisition environments~\cite{10449471}.
Research has largely been divided into two domains: enhancing attack potency and developing robust detection mechanisms. On the attack front, Timon~\cite{Timon_2019} demonstrated that CNNs can effectively perform non-profiled attacks, exploiting translation invariance to recover keys from de-synchronized traces where classical CPA fails. Similarly, Hasnain et al~\cite{10089540} established a framework for FPGAs that utilizes CNNs to profile voltage fluctuations, successfully extracting secret keys from shared power distribution networks. Addressing the data-dependency limitations of these profiled attacks, Yang et al~\cite{YANG2025100333} introduced Invariant Information Clustering (IIC-SCA), an unsupervised learning approach that maximizes mutual information to extract leakage without labeled datasets. Conversely, on the detection front, Bauer et al~\cite{10723110} proposed an ML-based monitor to identify physical tampering precursors to SCA, such as heat sink removal or ambient temperature manipulation, rather than analyzing the traces themselves. Our work bridges these domains by focusing on the real-time classification of the cryptographic workloads themselves, serving as a necessary context-aware trigger for the defensive or monitoring systems discussed in the literature.

\vspace{-0.8 em}
\subsection{Threat Model}
We consider a grey-box passive adversary with physical access to the target heterogeneous platform, capable of monitoring power consumption through non-invasive probes or shared voltage rails. We assume the system's logical security perimeters remain intact, preventing the attacker from inspecting memory or modifying the bitstream, thereby forcing reliance on physical side-channel leakage. The adversary's objective is to infer the execution state of the accelerator fabric by correlating instantaneous switching activity with specific workload signatures. In this setting, the attacker captures high-resolution power traces to distinguish between diverse computational tasks without disrupting the system's operational availability.

\vspace{-0.7 em}
\section{EXPERIMENTAL METHODOLOGY}
Building upon the leakage mechanisms and threat model discussed in Section II, this section describes the experimental pipeline used to capture, process, and classify real-time power side-channel traces from heterogeneous FPGA workloads. We describe the FPGA-based measurement setup, workload integration strategy, power trace acquisition process, dataset construction, and the hierarchical machine-learning architecture used for workload identification. The methodology is designed to evaluate both cryptographic and non-cryptographic accelerators under controlled and mixed-activity conditions, ensuring consistency across all evaluated workloads, including AES, ECC, SHA-256, CNN, and idle execution states.

\vspace{-0.7 em}
\subsection{FPGA Integration and Power Trace Acquisition}
In this work, we construct a complete FPGA-ChipWhisperer measurement pipeline to observe, record, and analyze side-channel leakage from heterogeneous cryptographic workloads. Figure~\ref{fig:dynsca_overview} illustrates the overall setup: the attacker (1) interacts with a host computer running Python scripts that configure the FPGA, schedule workloads, and capture power traces; the FPGA accelerator fabric (2) executes multiple instantiated hardware implementations, including AES, ECC, SHA-256, and a CNN accelerator, on the Artix-7 CW305 platform; and the ChipWhisperer capture board (3) performs synchronized power measurement of the FPGA core rail during execution. This environment enables deployment of both cryptographic and non-cryptographic accelerators, as well as controlled mixed-activity configurations where inactive modules contribute background leakage. The measurement workflow emulates an attacker ``sneaking into'' the FPGA execution path and observing instantaneous power signatures to infer the active workload.

\textbf{FPGA Bitstream Preparation:}
To enable controlled experimentation across different hardware workloads, we prepared a set of FPGA bit streams targeting the Xilinx Artix-7 A100T device on the CW305 evaluation platform. Each bitstream corresponds to a distinct hardware configuration, allowing isolation of individual accelerator power signatures as well as characterization of background leakage when multiple modules coexist on the same fabric. All designs were synthesized and implemented using Vivado, and the resulting configuration files were programmed onto the FPGA through the ChipWhisperer API, ensuring a consistent and reproducible deployment process across all experiments. The overall flow of bitstream generation to dataset construction is shown in Figure~\ref{fig:flow_pipeline}.

\begin{figure}[!t]
    \centering
    \includegraphics[width=0.8\linewidth]{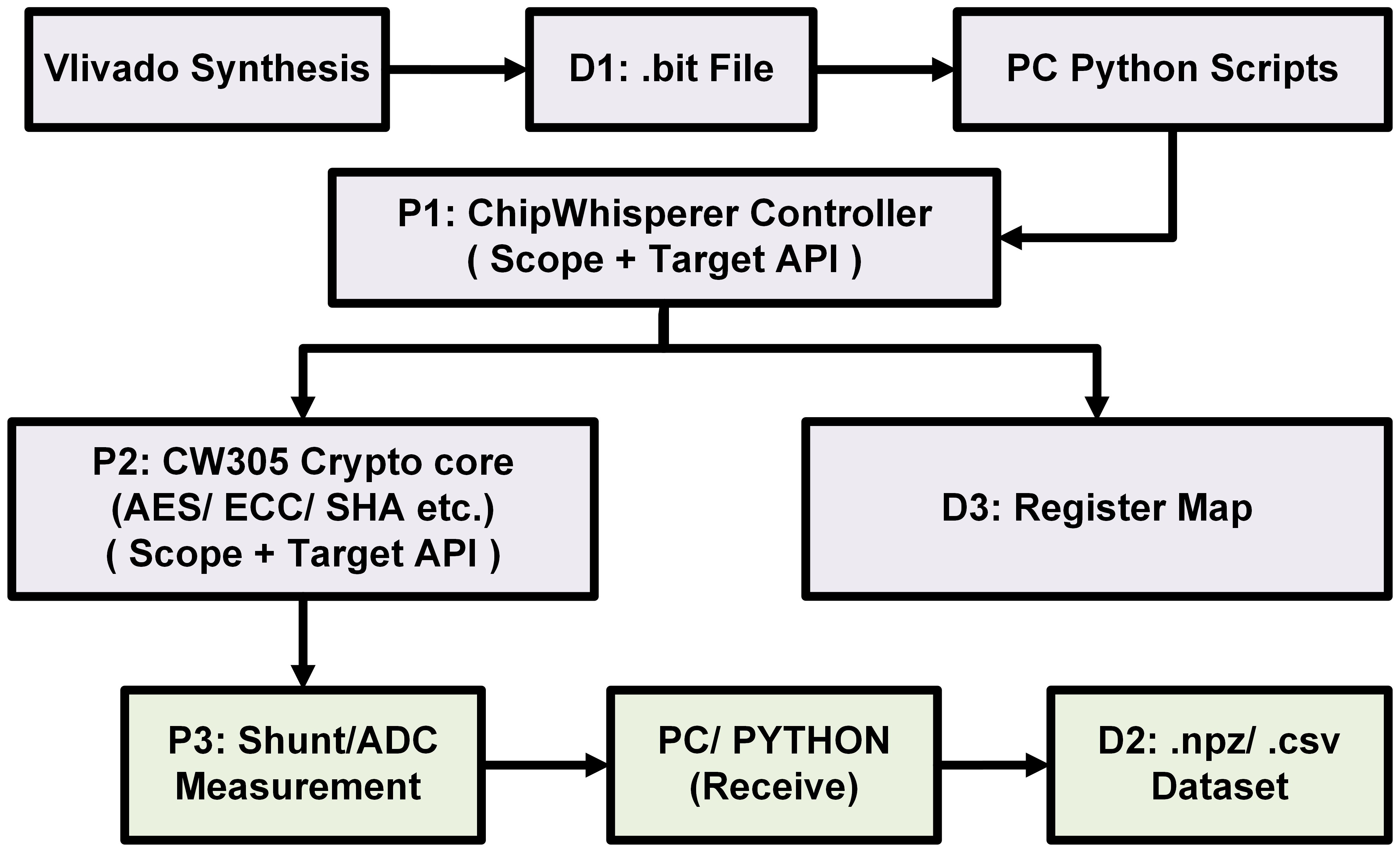}
    \caption{End-to-end measurement pipeline from bitstream generation to \texttt{.csv} dataset creation for ML prediction.}
    \label{fig:flow_pipeline}
    \vspace{-1.2 em}
\end{figure}

The first group of designs consisted of single-accelerator configurations. An AES-only bitstream instantiated a standalone AES-128 encryption core, providing a compact and fast workload with a short, burst-like power trace. In parallel, an ECC-only bitstream implemented a full P-256 scalar multiplication engine whose substantially longer execution time allows us to examine prolonged and high variance leakage patterns. We additionally generated an idle configuration in which no accelerator is triggered. This design captures the baseline dynamic and static leakage of the device and serves as a reference for distinguishing true cryptographic activity from background noise. We also incorporated a SHA-256 implementation sourced from the open-source Secworks repository~\cite{Secworks_sha256_2025} into an additional bitstream. This design introduces a third cryptographic workload with computational and leakage characteristics distinct from both AES and ECC. Finally, a CNN accelerator bitstream was integrated to represent a non-cryptographic, input-driven workload while enabling evaluation of power-side-channel behavior~\cite{Badhan2025CNNVerilog}. Together, these bitstreams form a comprehensive suite of hardware configurations that enable fine-grained exploration of power side-channel leakage across heterogeneous FPGA-based workloads.

\textbf{Runtime Mode Selection for Co-located Accelerators:}
To evaluate whether ML models can distinguish accelerators under realistic heterogeneous multi-tenant conditions, we combined AES and ECC accelerators into a single shared FPGA bitstream. This integrated design allows both accelerators to coexist on the Artix-7 fabric simultaneously, enabling analysis of how background static leakage and overlapping dynamic power contributions affect the resulting power traces. This selective activation is controlled through a mode-selection register mapped at address \texttt{0x0100}, which determines how the shared \texttt{crypt\_start} signal is routed within the hardware. When the register value is set to \texttt{0x01}, the start pulse is forwarded to the AES encryption engine while the ECC core remains clocked but computation-idle, producing only baseline leakage. Conversely, when the register holds \texttt{0x00}, the ECC scalar-multiplication engine receives the start trigger and executes its full computation while the AES unit remains inactive but instantiated. This runtime-switching mechanism eliminates the need to reprogram the FPGA between workloads and allows the system to emulate a heterogeneous execution environment in which multiple accelerators coexist but only one performs active computation at any given moment.

\begin{figure*}[!t]
\centering

\subfloat[AES power trace.]{%
  \includegraphics[width=0.48\linewidth]{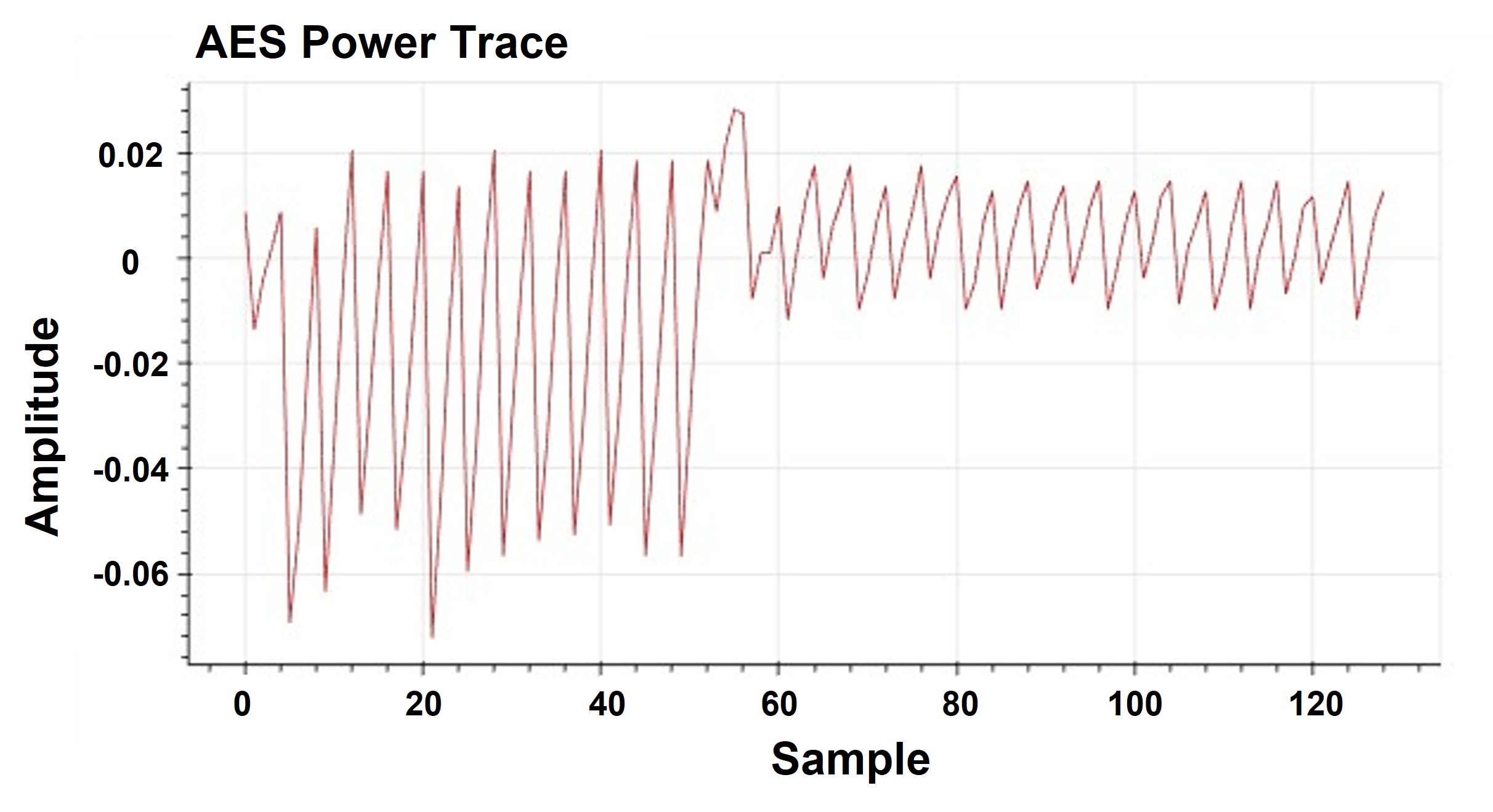}%
  \label{fig:trace_aes}%
}\hfill
\subfloat[ECC power trace.]{%
  \includegraphics[width=0.48\linewidth]{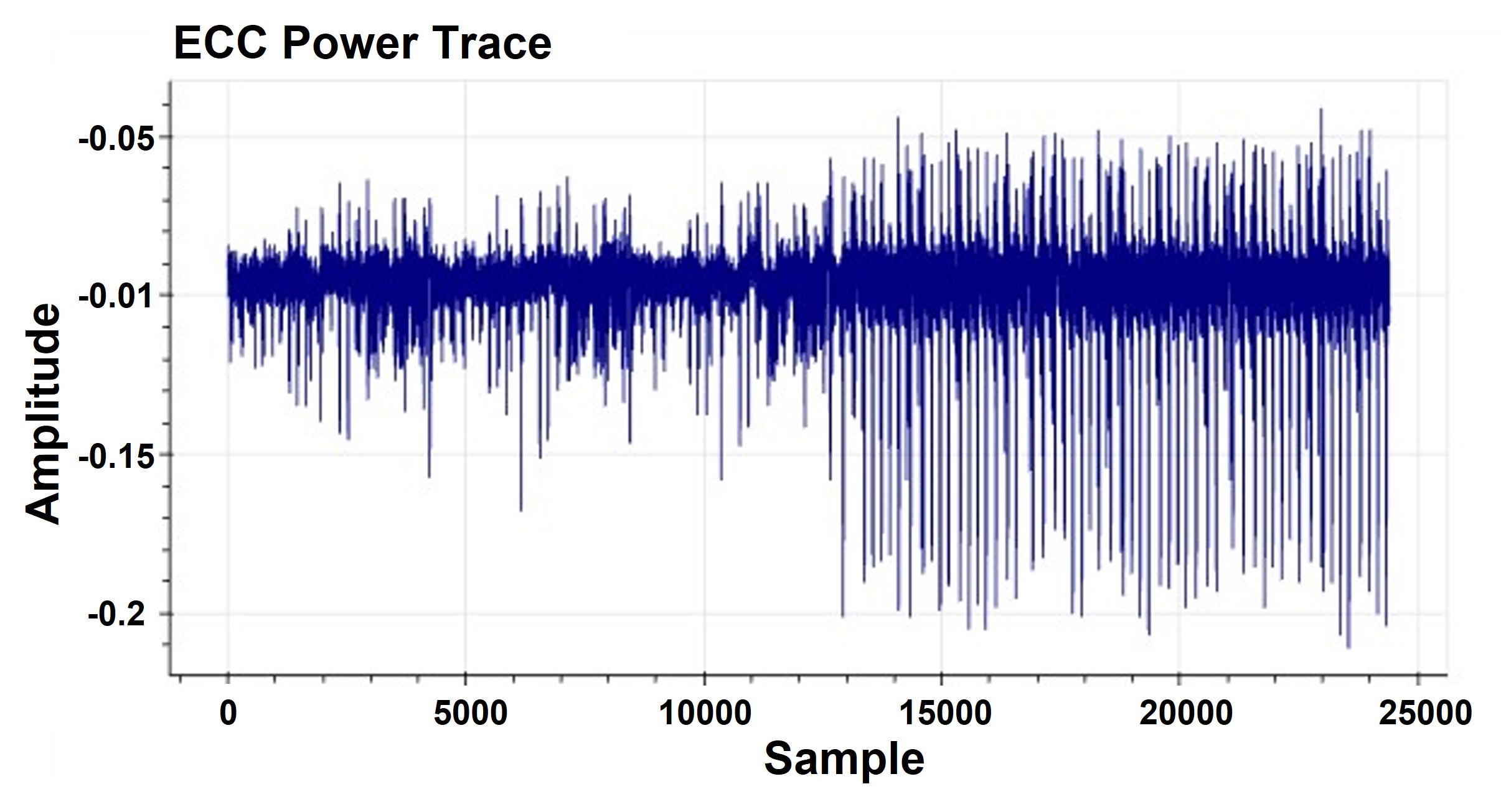}%
  \label{fig:trace_ecc}%
}

\subfloat[Idle FPGA power trace.]{%
  \includegraphics[width=0.48\linewidth]{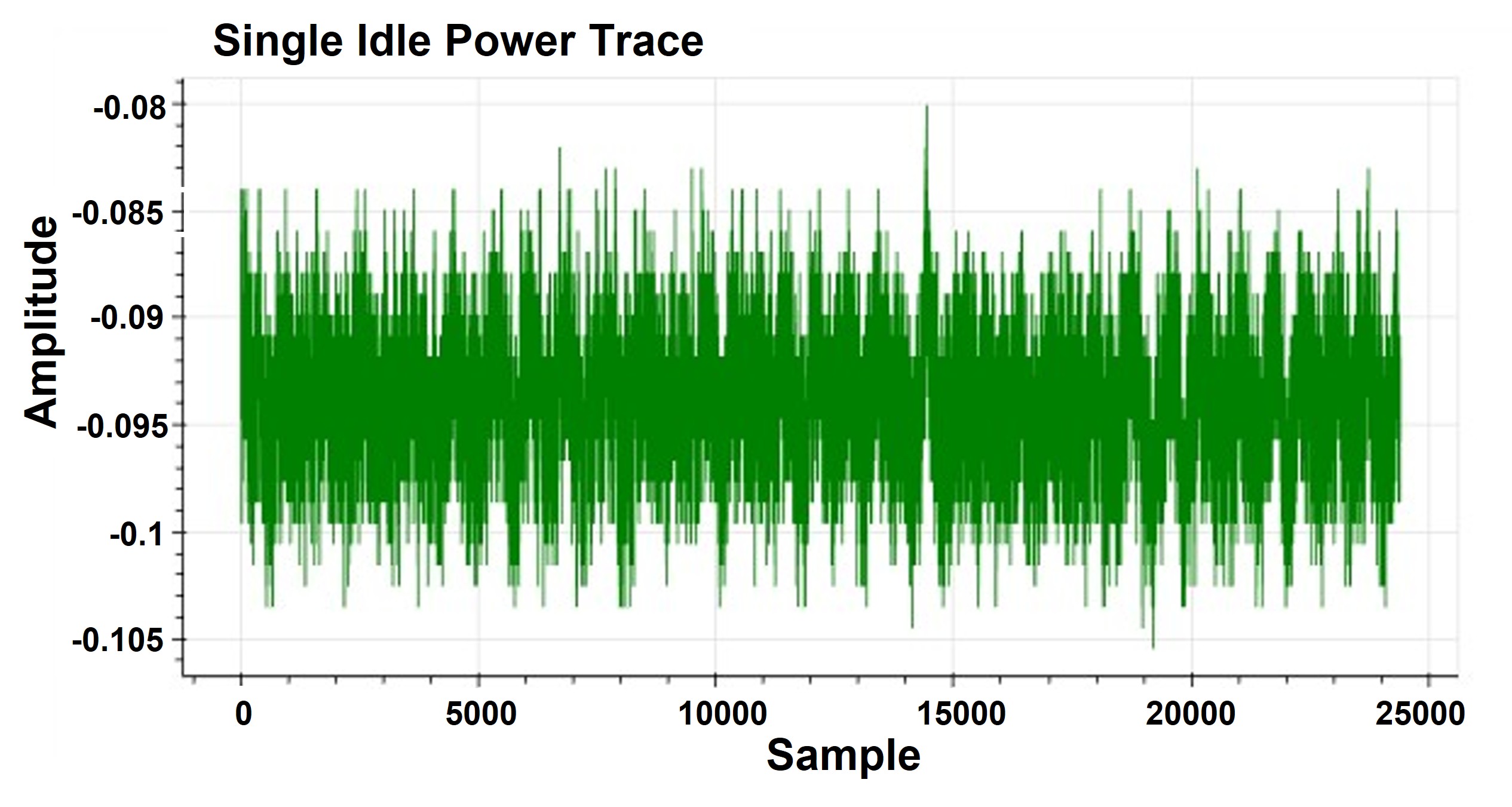}%
  \label{fig:trace_idle}%
}\hfill
\subfloat[SHA-256 power trace.]{%
  \includegraphics[width=0.48\linewidth]{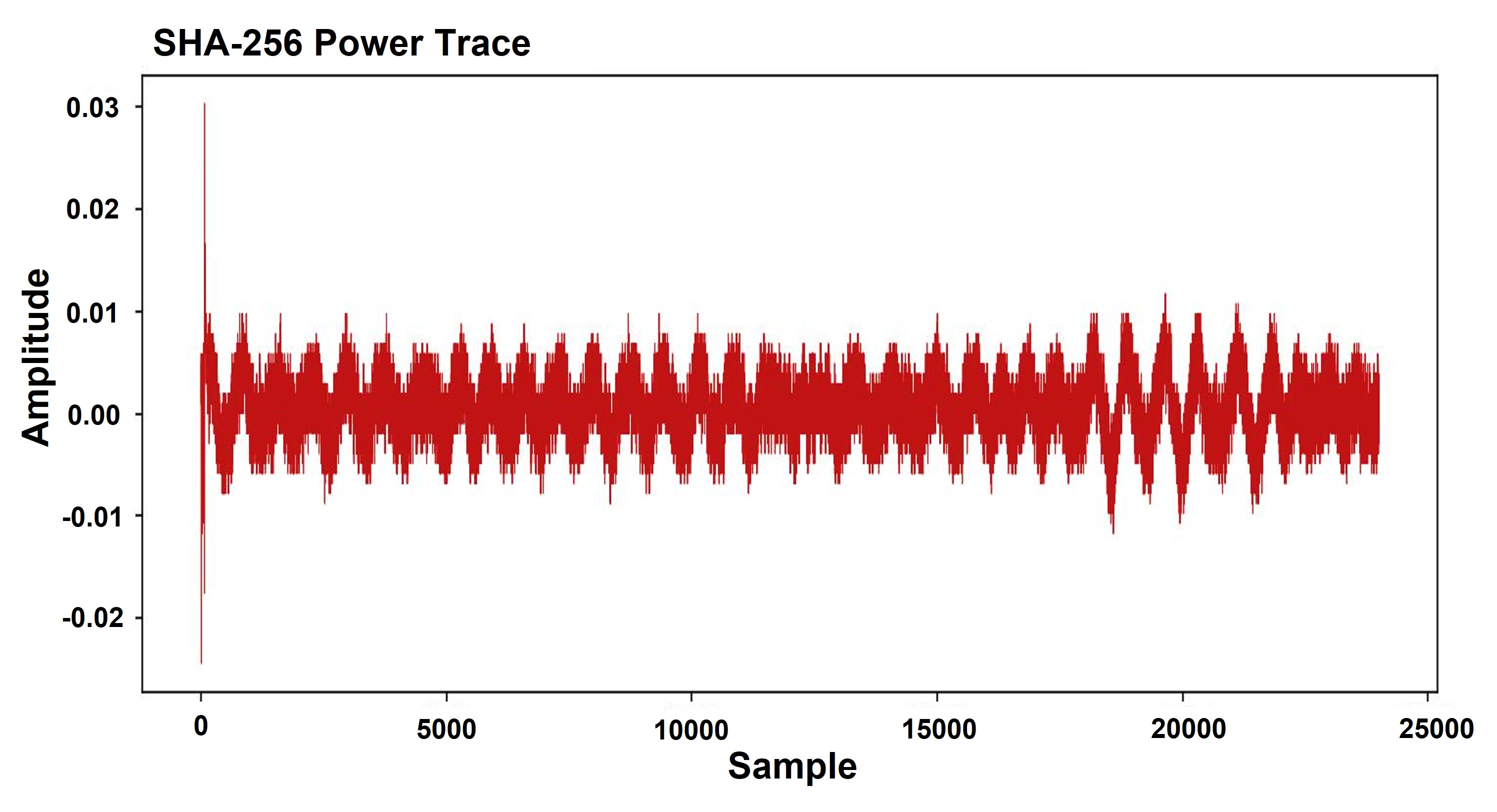}%
  \label{fig:trace_sha}%
}

\caption{Representative single power traces captured from the CW305 platform for (a) AES encryption, (b) ECC scalar multiplication, (c) an idle but clocked FPGA, and (d) SHA-256 hashing. The distinct trace lengths and shapes reflect the different execution latencies and switching characteristics of each workload.}
\label{fig:single_traces}
\vspace{-1.2 em}
\end{figure*}

\textbf{Power Measurement Setup:}
To capture fine-grained power signatures from FPGA-resident cryptographic and a CNN accelerator, we employ the ChipWhisperer-Lite measurement platform coupled with the CW305 Artix-7 evaluation board. The CW305 provides a dedicated shunt-resistor-based measurement path on the \textit{VCCINT} rail, enabling direct observation of instantaneous current draw from the programmable logic fabric. This hardware-level isolation allows us to bypass the noise, power gating, and dynamic voltage-scaling effects typically present in larger SoCs, ensuring that the recorded traces primarily reflect the switching activity of the targeted accelerator.

The ChipWhisperer scope is configured to use the CW305's TIO4 pin as a rising-edge trigger source, which is asserted automatically when the selected hardware accelerator core begins execution. Once armed, the scope samples the shunt voltage at a rate determined by the configured ADC window. Because each accelerator exhibits a different execution latency, we tailor the capture settings accordingly. For AES, which completes in only a few tens of clock cycles, we record a short window of 130 ADC samples with no decimation, fully covering the encryption burst. In contrast, ECC scalar multiplication and SHA-256 hashing run over several milliseconds, so we acquire 24,500 samples with a 100$\times$ decimation factor. The same long window is reused for the idle configuration so that background leakage is measured over a duration comparable to the active workloads. Across all workloads, these settings ensure that the recorded traces are adequately long to include the full computation interval while minimizing redundant samples. Representative single power traces for AES, ECC, idle, and SHA-256 are shown in Figure~\ref{fig:single_traces}.  This consistency is particularly important for downstream ML classification, where variations in sampling alignment or truncation can degrade model performance.

\textbf{Data Collection Workflow and Dataset Composition:}
The full data acquisition pipeline is orchestrated through Python scripts that interface with the CW305 via the ChipWhisperer API. For each workload, the corresponding input parameters are first written into the accelerator’s memory-mapped register space, including AES encryption keys and plaintext blocks as well as ECC scalars and elliptic-curve point coordinates.
Once configured, the script arms the ChipWhisperer scope, issues a \texttt{target.go()} command to trigger execution, and retrieves both the resulting ciphertext or output data and the corresponding power trace. Each captured waveform is immediately stored alongside its metadata, ensuring traceability of keys, mode settings, workload type, and timing information.

To support statistical analysis and ML-based classification, we construct large-scale datasets under controlled input conditions. Across all cryptographic workloads evaluated in this study, we adopt a consistent data collection methodology in which each application is measured over 10,000 power traces organized into ten distinct input groups. For AES, each group corresponds to a fixed key-plaintext pair, enabling observation of both intra-group similarity and inter-group leakage variations. ECC follows the same structure using fixed scalar-point pairs. Using the combined AES-ECC bitstream, we additionally record mixed-mode traces in which only one accelerator is active while the other remains idle, covering both the AES-active/ECC-idle and ECC-active/AES-idle scenarios. A separate SHA-256 dataset is obtained under the same collection policy, using ten fixed message block inputs to generate another 10,000 traces. All 10,000 traces for each workload are exported to \texttt{.csv} format and serve as the primary training corpus for our ML model. After training, we generate a separate evaluation set consisting of 100 additional traces formed from ten new input cases per workload and store them in an independent \texttt{.csv} file to ensure that model testing is performed on previously unseen data.

To broaden the evaluation scope beyond cryptographic primitives, we developed a dedicated measurement workflow for a CNN accelerator from the base hardware design~\cite{Badhan2025CNNVerilog}. The custom Verilog core, deployed on the CW305 Artix-7 fabric, processes $16\times 16~\text{pixel}$ inputs through a fixed multi-stage sequential processing architecture consisting of Laplacian-based filtering, convolutional processing, $2\times 2$ max pooling, and threshold-based output generation. Unlike cryptographic cores, where power leakage is dominated by key-dependent control flow and round structure, the CNN's power consumption is primarily driven by data-dependent switching activity within its convolutional and pooling logic. To capture this variability while preserving a single workload class, we acquire 10,000 traces, which means 2,500 per class of input across four fixed input topologies, including uniform gray, horizontal gradient, checkerboard, and high-contrast corner blocks. This diversity ensures that the learned signature reflects the accelerator's execution behavior rather than specific image content.

Physical analysis of the captured traces reveals leakage characteristics correlated with input complexity. As shown in Figure~\ref{fig:cnn_avg_traces}, the plotted traces correspond to voltage measurements captured on the FPGA VCCINT rail, which are proportional to instantaneous power consumption but are shown in arbitrary units due to the absence of absolute power calibration. Consequently, the analysis focuses on relative amplitude variations and temporal structure, rather than absolute power values. Specifically, the 5,000-sample execution window exhibits amplitude variation distributed throughout the active processing stages. Inputs with higher spatial frequency, such as checkerboard patterns, induce increased switching activity within the datapath registers, resulting in subtle amplitude variations compared to uniform inputs, as visible in the averaged traces. The trace characteristics reflect the cumulative power signature of the state machine progressing through load, convolution, pooling, and threshold stages. The hierarchical classifier exploits these data-dependent transient features to robustly distinguish the CNN workload from cryptographic and idle execution. A separate validation set of 100 traces, generated from new instances of the same input patterns, is used to evaluate generalization and reduce overfitting.

\vspace{-1 em}
\subsection{Data Acquisition and Preprocessing}
To robustly classify cryptographic/non-cryptographic workloads and idle states from power side-channel leakage traces, a hierarchical machine learning framework was developed utilizing a two-stage Random Forest classifier. While recent literature prioritizes Deep Learning architectures for side-channel analysis, we selected the Random Forest algorithm to prioritize model interpretability and computational efficiency. Unlike neural networks, this approach offers intrinsic feature importance metrics and allows for the explicit validation of physical leakage characteristics while requiring significantly lower training complexity.  This makes it highly suitable for resource-constrained embedded monitoring. The methodology employs a hybrid data preprocessing strategy to accommodate varying dataset structures and leverages the two-stage hierarchy to effectively distinguish between short-duration, high-frequency operations (AES) and longer, distinct workloads (ECC, SHA-256, CNN, Idle).
The experimental dataset comprises power traces captured from an Artix-7 FPGA (CW305) performing five distinct classes of operations: AES-128, ECC, SHA-256, a Convolutional Neural Network (CNN) accelerator, and System Idle. A hybrid partitioning strategy was implemented to address heterogeneity in data availability and format:
\begin{itemize}
    \item \textbf{Standard Classes (AES, ECC, SHA-256, Idle):}
    For these classes, distinct datasets were utilized for training and testing to simulate realistic deployment scenarios where the model encounters unseen data. Training was performed on large repositories (approx. 10,000 traces per class), while evaluation was conducted on separate, isolated validation files (approx. 100 traces each).
    \item \textbf{CNN Workload:}
    Due to data constraints, the CNN dataset (10,000 traces) was processed using a random stratified split, designating the majority for training and reserving a fixed subset (100 traces) for testing.
    \item \textbf{Mixed Traces:}
    To enhance model robustness against background noise and interleaved operations, the training was augmented with existing mixed-signal traces. These traces feature active AES operations occurring alongside Idle states and active ECC operations, ensuring the model learns to isolate specific cryptographic signatures within noisy environments.
\end{itemize}

\begin{figure}[!t]
    \centering
    \includegraphics[width=\linewidth]{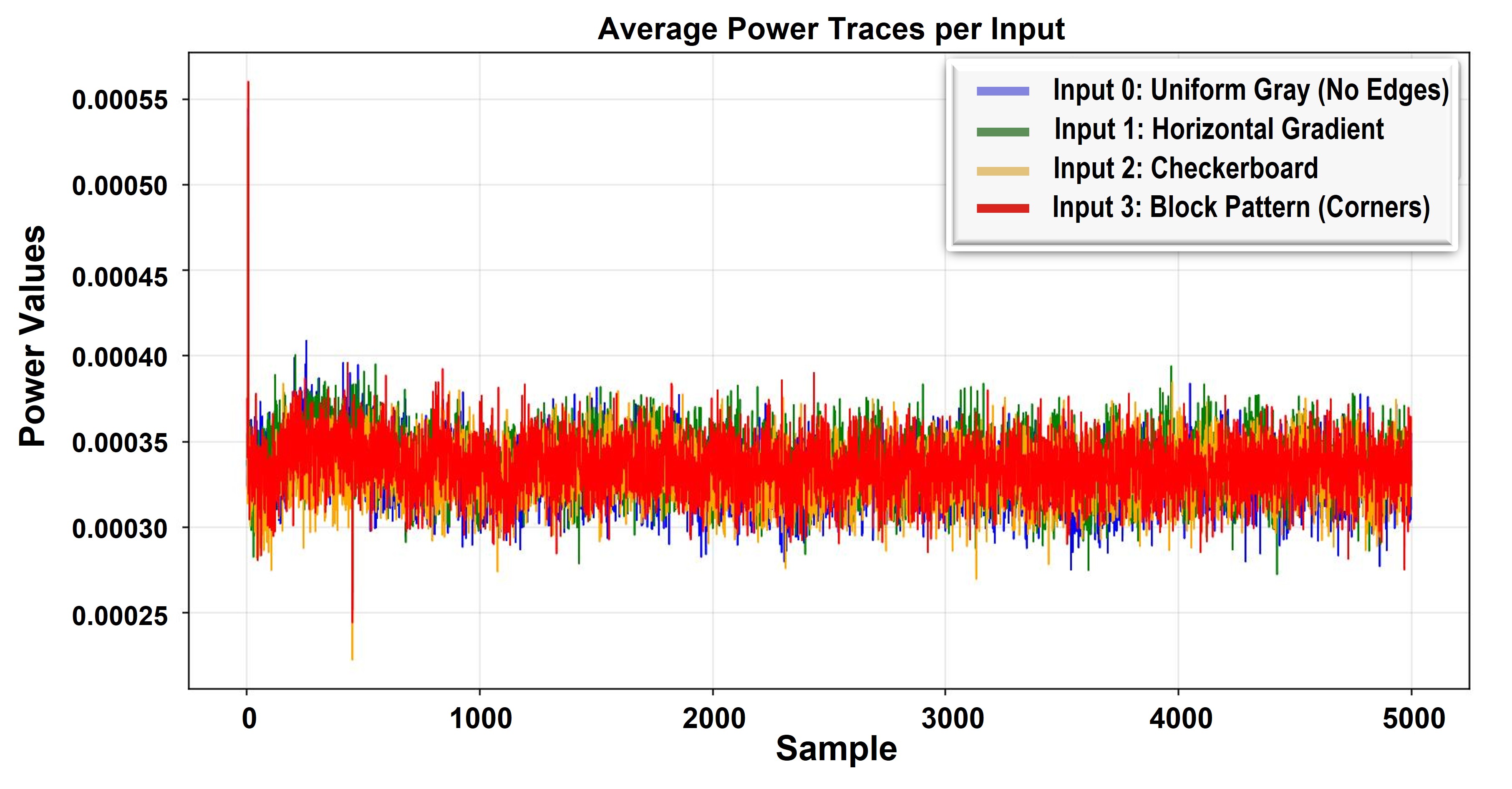}
    \caption{Average CW305 VCCINT power traces for the CNN accelerator under four fixed input patterns, highlighting data-dependent switching during convolution-dominated execution stages.}
    \label{fig:cnn_avg_traces}
\end{figure}

\vspace{-0.9 em}
\subsection{Feature Extraction}
Raw power traces were first subjected to zero-mean normalization to remove DC offset variations. To convert the variable-length time-series data into fixed-length input vectors, a set of 21 statistical and spectral features was extracted for each trace.
\begin{itemize}
    \item \textbf{Time-Domain Features:}
    Includes Mean, Standard Deviation, Root Mean Square Error (RMSE), Energy, Skewness, Kurtosis, and Entropy. Advanced shape descriptors such as Crest Factor, Impulse Factor, and Shape Factor were also calculated to capture the unique vertical asymmetries of the power spikes.
    \item \textbf{Frequency-Domain Features:}
    A Fast Fourier Transform (FFT) was applied to extract the Power Spectrum Peak, Spectral Centroid, and Spectral Roll-off. These features are critical for distinguishing between hardware cores operating at different switching frequencies or intensities.
\end{itemize}
\begin{figure}[!h]
\centering
\vspace{-1em}
\includegraphics[width=0.8\linewidth]{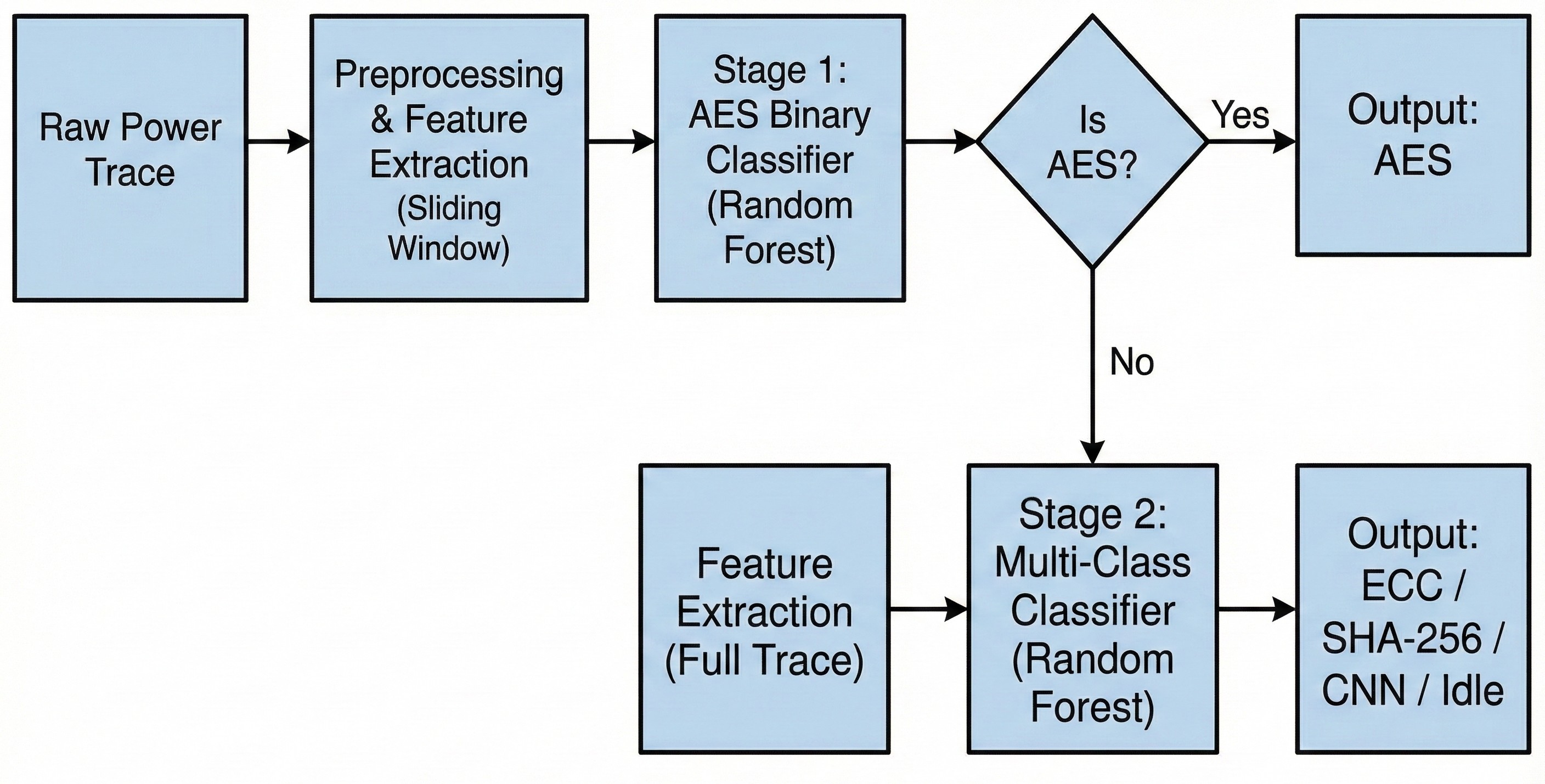}
\caption{Block diagram of the hierarchical classifier model over all power traces.}
\label{fig:block}
\end{figure}

\begin{figure}[!b]
\centering
\includegraphics[width=0.75\linewidth]{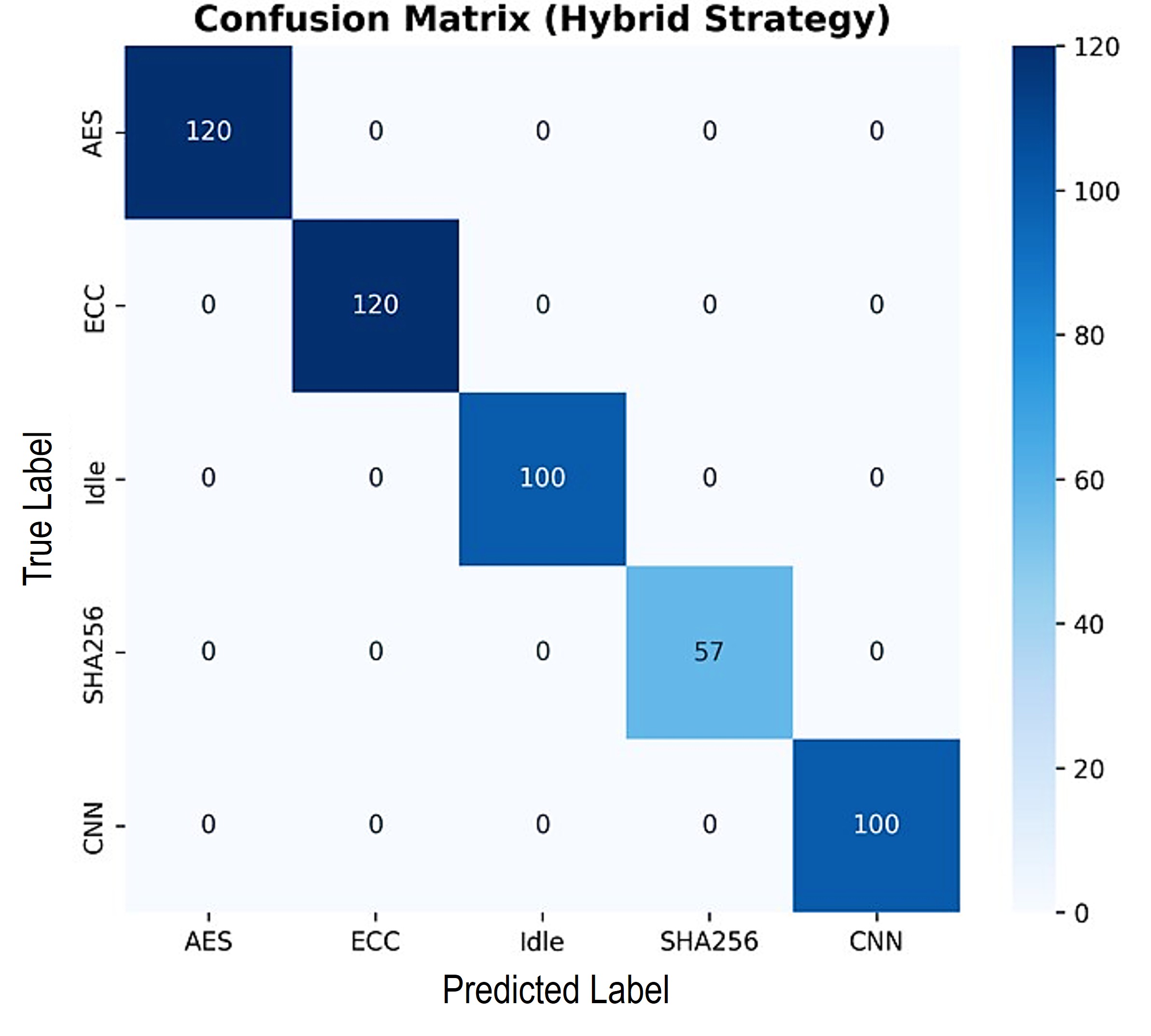}
\caption{Confusion matrix of classifier prediction versus actual class to show how accuracy of classifier prediction over the testing dataset.}
\label{fig:matrix}
\vspace{-1.5em}
\end{figure}

\begin{table}[!t]
\caption{Summary of Extracted Features for Power Trace Classification}
\begin{center}
\begin{tabularx}{\columnwidth}{|c|l|X|}
\hline
\textbf{Domain} & \textbf{Group} & \textbf{Specific Metrics} \\
\hline
\multirow{3}{*}{Time} & Stat. & Mean, Std Dev, Skewness, Kurtosis, Entropy \\
\cline{2-3} 
 & Amp. & Energy, RMSE, Peak-to-Peak Ratio \\
\cline{2-3} 
 & Shape & Crest, Impulse, Shape, Clearance Factors \\
\hline
\multirow{2}{*}{Freq.} & Spectral & Power Spectrum Peak, Spectral Centroid \\
\cline{2-3} 
 & Band. & Spectral Rolloff (85\% Energy) \\
\hline
\end{tabularx}
\label{tab:features}
\vspace{-1.5 em}
\end{center}
\end{table}

\vspace{-0.9 em}
\subsection{Hierarchical Classification Architecture}
A standard multi-class classifier often struggles to distinguish between short, bursty operations, such as AES, and long-running processes, such as ECC or CNN, simultaneously. To resolve this, a Hierarchical Random Forest architecture was implemented Figure~\ref{fig:block}.

\begin{figure*}[!htbp]
\centering
\includegraphics[width=0.75\textwidth]{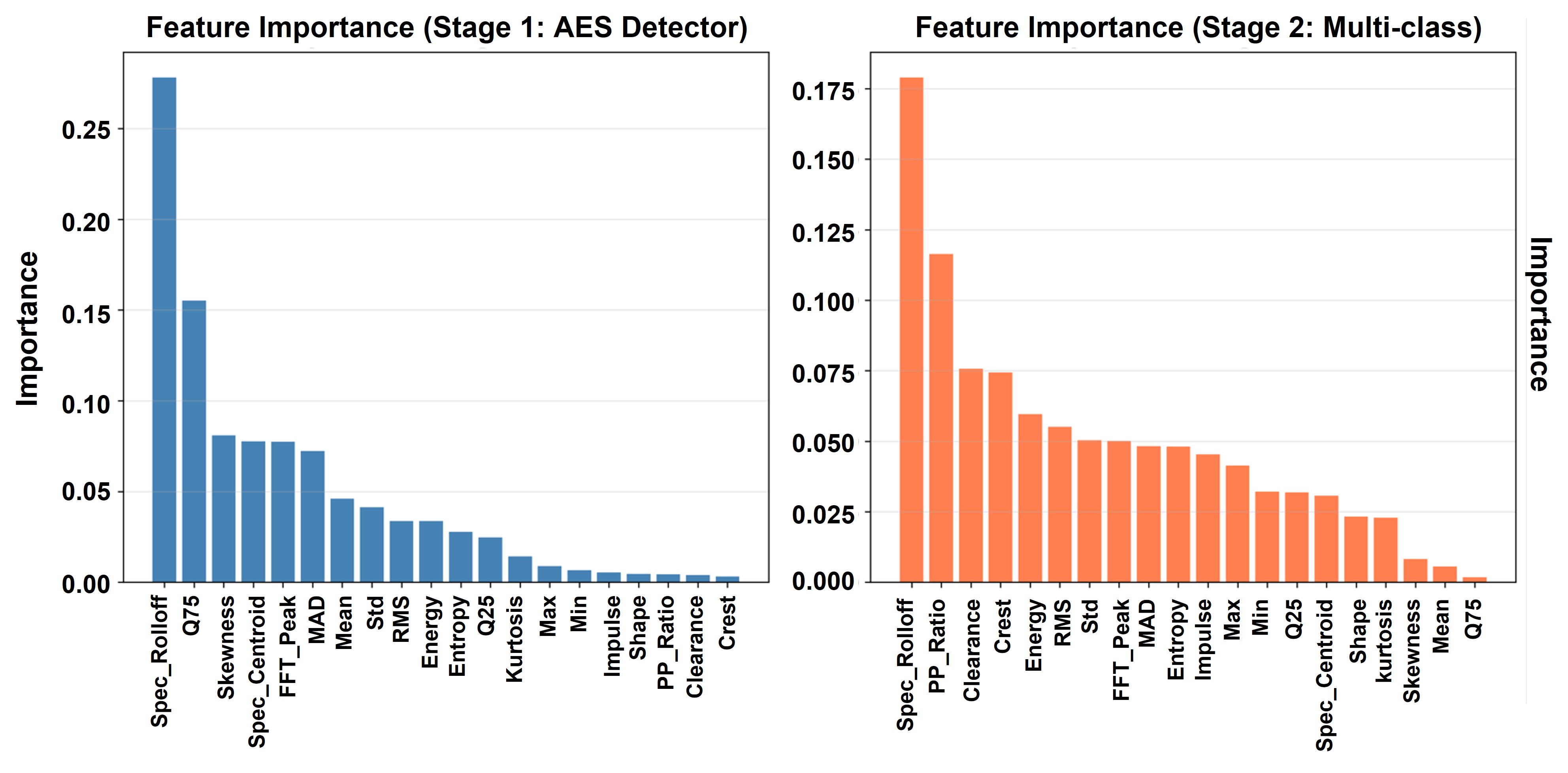}
\caption{Gini impurity reduction illustrating feature importance across hierarchical classifier stages.}
\label{fig:gini}
\end{figure*}

\textbf{Stage 1. Binary AES Detector (Sliding Window):} The first stage is a binary classifier trained to detect ``AES'' vs. ``Rest.'' Since AES operations are short and distinct, this stage utilizes a sliding window approach (window size = 129 samples, stride = 32). If any window within a trace yields an AES probability exceeding a threshold ($\tau >$ 0.45), the entire trace is classified as AES. This allows the model to pinpoint AES leakage even if it is temporally shifted.

\textbf{Stage 2. Multi-Class Classifier:} If Stage 1 is negative, the trace is passed to Stage 2. This classifier utilizes features extracted from the full trace duration to discriminate between the remaining classes: ECC, Idle, SHA-256, and CNN.
Both stages utilize Random Forest estimators (N=100) with balanced class weights to handle potential dataset imbalances.


\section{EXPERIMENTAL RESULTS}
The proposed hybrid hierarchical model was evaluated on the unseen test partitions described in Experimental Methodology. The performance metrics focus on classification accuracy, confusion matrix analysis, and feature relevance.

\vspace{-0.7 em}
\subsection{Classification Performance} 
The model achieved a global accuracy of 100\% across all test scenarios, effectively distinguishing between the five target classes and the mixed-signal environments. As shown in the confusion matrix in Figure~\ref{fig:matrix}, the model produced no false positives or false negatives on the test set.

\begin{itemize}
        \item \textbf{Standard Workloads:} The model correctly identified all discrete AES, ECC, Idle, SHA-256, and CNN traces.
        \item \textbf{Mixed Workloads:} Crucially, the model maintained 100\% accuracy on the ``AES Mixed'' and ``ECC Mixed'' datasets. This validates the efficacy of the sliding window approach in Stage 1, which successfully isolated AES signatures even when interleaved with other system activities.
\end{itemize}

\begin{figure}[b!]
\centering
\includegraphics[width=0.8\linewidth]{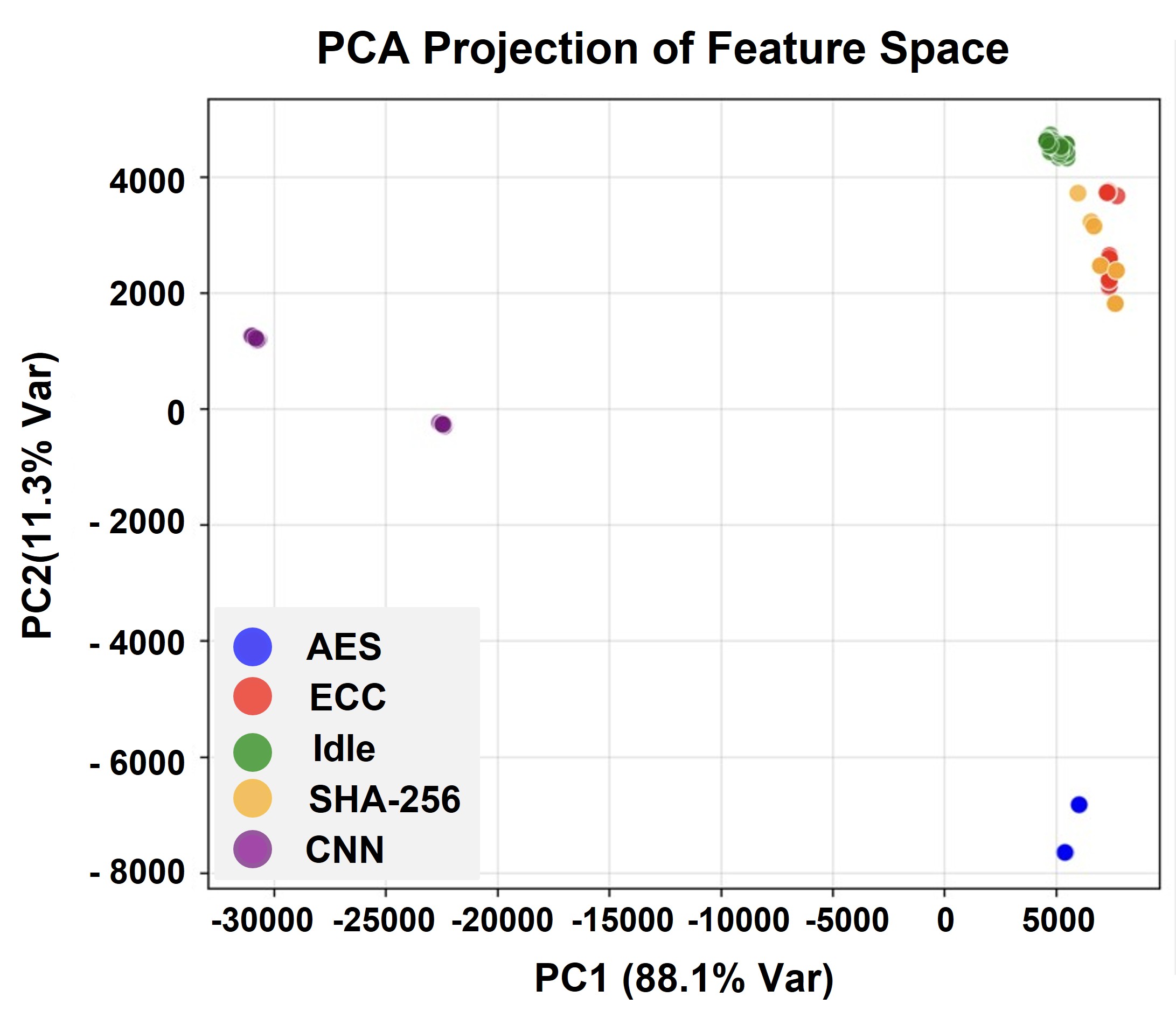}
\caption{Principal Component Analysis projection of feature space to show off distinct clustering of features with each class of power trace. Note the distinct clustering of CNN and Idle states, and the proximity of AES and ECC, which is resolved by the hierarchical classifier.}
\label{fig:pca}
\vspace{-1em}
\end{figure}

\vspace{-0.7 em}
\subsection{Feature Importance Analysis:} 
An analysis of Gini impurity reduction revealed distinct feature hierarchies for the two stages, as illustrated in Figure~\ref{fig:gini}

\begin{itemize}
    \item \textbf{Stage 1 (AES vs. Rest):} The Spectral Rolloff and Spectral Centroid were the most dominant features. This suggests that the high-frequency switching characteristics of the AES hardware implementation are its most distinguishing factor against other workloads.
    \item \textbf{Stage 2 (Multi-Class):} While spectral features remained important, time-domain metrics such as Peak-to-Peak Ratio and Crest Factor gained significance. This indicates that distinguishing between ECC, SHA-256, and CNN relies more heavily on the amplitude profile and “shape” of the power envelope over longer durations.
\end{itemize}

\vspace{-0.7 em}
\subsection{Feature Space Visualization:} To validate the discriminative power of the selected feature set, Principal Component Analysis (PCA) was applied to project the 21-dimensional feature vectors into a 2-dimensional latent space, as illustrated in Figure~\ref{fig:pca}.

The projection reveals well-defined, compact clusters for the CNN (purple), SHA-256 (orange), and Idle (green) classes, confirming that their power signatures are highly distinct in the chosen feature domain. Notably, the AES (blue) and ECC (red) clusters exhibit closer proximity in this low-dimensional projection. This proximity physically justifies the necessity of the proposed hierarchical architecture. In particular, by isolating AES in the first stage using specialized spectral features, the model effectively resolves the potential ambiguity between AES and ECC that might confound a simple flat classifier. The tight clustering of the CNN traces further confirms that the neural network accelerator exhibits a stable and unique power signature compared to the varying cryptographic operations.
\vspace{-0.7 em}

\section{FUTURE DIRECTIONS}
While the proposed hierarchical model achieved 100\% accuracy on the test dataset, this perfect performance raises a valid concern regarding potential overfitting to the specific noise profiles and experimental conditions of the controlled dataset. Such high precision often indicates that a supervised model has captured the exact distribution of the training data, which may not generalize well to the dynamic variances found in real-world deployments. These real-world deployments on reconfigurable heterogeneous architectures introduce a significant challenge to model robustness and reliability. Specifically, the IIC-SCA method proposed by Yang et al.~\cite{YANG2025100333} demonstrates the viability of unsupervised clustering for leakage detection without reliance on labeled ground truth. We plan to explore beyond these findings to ensure the system can adapt to the diverse and shifting operational profiles without the risk of overfitting to a static training set. Future work will also focus on expanding the experimental scope to include a wider variety of mixed-signal traces and additional application classes to ensure robustness against highly heterogeneous FPGA workloads.  Additionally, we intend to explore lightweight 1D-CNN architectures to address complex environmental distortions. As noted by Timon~\cite{Timon_2019}, such models possess inherent translation invariance that could improve resilience against clock jitter and temporal misalignment. 
\vspace{-0.7 em}
\section{CONCLUSION}
This work presented HAWC-Leak, a hierarchical machine learning framework capable of fingerprinting heterogeneous accelerator workloads with 100\% accuracy using non-invasive power side-channel analysis. Using fine-grain power capture on the CW305 platform, we showed that distinct accelerator workloads exhibit identifiable power signatures even in the presence of background leakage from idle modules, validating power as a practical side-channel for runtime inference. While the current framework supports only a single active workload at a time, the results establish a strong baseline for scalable workload fingerprinting.
\vspace{-0.7 em}

\bibliographystyle{IEEEtran}
\bibliography{ref}

\vfill

\end{document}